\documentclass[aps,twocolumn,floatfix,superscriptaddress,showkeys]{revtex4-2}

\usepackage{graphicx}
\usepackage{amsmath,amsfonts,amssymb,bm}

\usepackage{lmodern}
\usepackage{xcolor}
\usepackage[normalem]{ulem}
\usepackage{float}
\usepackage{braket}
\usepackage{dsfont}
\usepackage{multirow}
\usepackage{booktabs}

\usepackage{algorithm}
\usepackage{algorithmic}

\usepackage{tikz}
\usepackage{mathdots}
\usepackage{yhmath}
\usepackage{cancel}
\usepackage{color}
\usepackage{siunitx}
\usepackage{array}
\usepackage{multirow}
\usepackage{amssymb}
\usepackage{gensymb}
\usepackage{tabularx}
\usepackage{extarrows}
\usepackage{booktabs}
\usetikzlibrary{patterns}
\usetikzlibrary{shapes}

\usepackage{subfigure}
\usepackage{subfigmat}
\usepackage{comment}

\usepackage[colorlinks,
            linkcolor=blue,    
            citecolor=blue,  
            urlcolor=blue,
 	        bookmarks=true,        
	        bookmarksopen=true,    
	        bookmarksnumbered=true,
]{hyperref}

\newcommand{\beq}{\begin{equation}}
\newcommand{\eeq}{\end{equation}}

\definecolor{magenta}{rgb}{1.0, 0.0, 1.0}

\usepackage{environ}         
\usepackage{etoolbox}        
\usepackage{graphicx}        

\newlength{\myl}
\let\origequation=\equation
\let\origendequation=\endequation

\RenewEnviron{equation}{
  \settowidth{\myl}{$\BODY$}                       
  \origequation
  \ifdimcomp{\the\linewidth}{>}{\the\myl}
  {\ensuremath{\BODY}}                             
  {\resizebox{\linewidth}{!}{\ensuremath{\BODY}}}  
  \origendequation
}

\usepackage{geometry}
\begin{document}
\title{Tricriticality in 4D U(1) Lattice Gauge Theory}

\author{Rafael C. Torres}

\affiliation{CeFEMA, Instituto Superior T\'{e}cnico, Universidade de Lisboa, Av. Rovisco Pais, 1049-001 Lisboa, Portugal}

\author{Nuno Cardoso}

\affiliation{CeFEMA, Instituto Superior T\'{e}cnico, Universidade de Lisboa, Av. Rovisco Pais, 1049-001 Lisboa, Portugal}

\author{Pedro Bicudo}

\affiliation{CeFEMA, Instituto Superior T\'{e}cnico, Universidade de Lisboa, Av. Rovisco Pais, 1049-001 Lisboa, Portugal}

\author{Pedro Ribeiro}

\affiliation{CeFEMA, Instituto Superior T\'{e}cnico, Universidade de Lisboa, Av. Rovisco Pais, 1049-001 Lisboa, Portugal}

\affiliation{Beijing Computational Science Research Center, Beijing 100084, China}

\author{Paul McClarty}
\affiliation{Laboratoire L\'{e}on Brillouin, CEA, CNRS, Universit\'{e} Paris-Saclay, CEA Saclay, 91191 Gif-sur-Yvette, France}
\affiliation{Max Planck Institute for the Physics of Complex Systems, Nöthnitzer Str. 38, 01187 Dresden, Germany}

\begin{abstract}
The 4D compact U(1) gauge theory has a well-established phase transition between a confining and a Coulomb phase. In this paper, we revisit this model using state-of-the-art Monte Carlo simulations on anisotropic lattices. We map out the coupling-temperature phase diagram, and determine the location of the tricritical point, $T/K_0 \simeq 0.19$, below which the first-order transition is observed. We find the critical exponents of the high-temperature second-order transition to be compatible with those of the 3-dimensional $O(2)$ model.
Our results at higher temperatures can be compared with literature results and are consistent with them.
Surprisingly, below $T/K_0 \simeq 0.05$ we find strong indications of a second tricritical point where the first-order transition becomes continuous. 
These results suggest an unexpected second-order phase transition extending down to zero temperature, contrary to the prevailing consensus. If confirmed, these findings reopen the question of the detailed characterization of the transition including a suitable field theory description.
\end{abstract}

\maketitle

\section{Introduction}
\label{sec:intro}

Gauge theories formulated on the lattice \cite{PhysRevD.10.2445} are central to our understanding of various nonperturbative phenomena such as confinement of color charges in quantum chromodynamics. They are also important in condensed matter physics as emergent low energy theories of interacting spins or electrons in crystalline solids or in quantum simulators. Pure U(1) gauge theory has a long history as a relatively simple example of a theory exhibiting charge confinement at least at strong coupling \cite{BANKS1977493, Fradkin1978, Bhanot1982, Moriarty1982, Barber1984, PhysRevD.19.619, AZCOITI1991207, AZCOITI1991, Arnold2003}. In three dimensions, Polyakov observed that the theory is confining at all couplings as a consequence of monopole condensation where the existence of monopole charges is a consequence of the compactness of the gauge group \cite{Polyakov:1987ez,POLYAKOV197582}. In 4D, the analogous excitations can be interpreted as monopole worldlines. Polyakov \cite{POLYAKOV197582} conjectured and Alan Guth proved \cite{Guth1980} that there is a phase transition between a strong coupling confined phase and a weak coupling Coulomb phase with propagating photon excitations. 

Evidence for this phase transition was found already in Monte Carlo simulations of the lattice action from the 1980s \cite{Creutz1979, Lautrup1980, Guth1980, Toussaint1981,Bhanot1981,MUTTER1982,Jersak1983,GUPTA198686, Evertz1985} and corroborated by later studies \cite{AZCOITI1991207, AZCOITI1991, Vettorazzo2004,Vettorazzo2004_2, Arnold2003, Berg2006, PhysRevD.88.065025}. An early controversy over the nature of the phase transition at zero temperature was eventually settled in favor of a weak first order transition \cite{Bhanot1981,MUTTER1982, Moriarty1982, Jersak1983,GUPTA198686,Evertz1985, AZCOITI1991, Lang1994, Arnold1999,COX1999,Campos_1998,CAMPOS1998b,KlausRoiesnel1998,ARNOLD2003864, Arnold2003, Vettorazzo2004,Vettorazzo2004_2,PhysRevD.88.065025, Berg2006}.

Monte Carlo results are extracted from finite size scaling of simulations performed on lattices of fixed temporal size $N_t$ and spatial extent $N_s$. Finite temperatures in this theory have been partially explored by simulating lattices with fixed $N_t$ while varying  $N_s$  \cite{Vettorazzo2004_2, Berg2006, PhysRevD.88.065025}. For $N_t = 1$, the 4D U(1) lattice gauge theory (LGT) is decoupled into a 3D U(1) LGT and a 3D XY model. While the 3D U(1) LGT is known to remain in the confining phase throughout the parameter space, the 3D XY model has a second order phase transition at a finite value of the coupling. As such, one expects a phase boundary connecting the second order transition at $N_t = 1$ to the first order transition at $N_t \rightarrow \infty$, along which there must be a tricritical point where the order of the phase transition changes. Evidence for this feature of the phase boundary at finite temperature has been seen numerically \cite{PhysRevD.88.065025}.   

Further studies considered corrections to the action in the form of the leading higher harmonic of the plaquette term~\cite{COX1999, Campos_1998, Cox:1998ji}. 
These corrections make it possible to shift the position of this tricritical point~\cite{COX1999, Campos_1998}. By tuning the ratio of couplings it is possible to shift the tricritical point to $T=0$, recovering a second order transition down to zero temperature~\cite{Cox:1998ji}.

In this paper, we revisit both the nature of the low temperature transition and the location of tricritical points by providing a unified phase diagram as a function of coupling and temperature for the original Wilson action.  It is natural to approach this problem from the perspective of a Hamiltonian formulation of the theory. However, it is the lattice action that can be simulated efficiently using quantum Monte Carlo. Furthermore, the standard lattice action for attainable system sizes provides access only to a very coarse-grained set of temperatures. 
We provide a systematic Monte Carlo study of the 4D pure U(1) gauge theory addressing both of these issues. 

The problem of sweeping in temperature is solved by introducing a continuous anisotropy parameter between the spatial and temporal directions on the lattice. 
Technically the parallel between the action and the Hamiltonian formulation can be drawn by making use of a Villain approximation at the expense of introducing a renormalization of couplings.
To fix this issue, we express all Hamiltonian quantities in terms of the critical coupling at zero temperature. In this way, we are able to systematically relate various anisotropic lattice sizes and couplings of the simulated action to the Hamiltonian couplings and temperature. Using these technical innovations in tandem, we were able to access the finite temperature phase diagram over a wider region of parameter space than hitherto explored. We find direct evidence for a second order transition at higher temperatures with criticality consistent with the expected 3D XY universality class. As the temperature is lowered and the transition moves to stronger coupling, we pinpoint the location of the tricritical point hinted at in previous studies. 
At lower temperatures still the first order transition becomes weaker and in the zero temperature limit our results are compatible both with a continuous and a very weakly first order transition. 
Although our results are not precise enough, they suggest
a scenario where the zero-temperature transition is continuous concomitant with the existence of an additional tricritical point at low temperature.  

The paper is organised as follows. In section \ref{sec:lgt}, we introduce the model and observables, the simulation procedure and lattice anisotropy parameter and the connection to temperature. Then, in section \ref{sec:results}, we outline how we identify the phase transition and present the complete finite temperature phase diagram of the 4D U(1) LGT. In section \ref{sec:opt}, we describe how to determine the order of the transition and present our results revealing a tricritical point. We discuss critical exponents in section~\ref{sec:exp} and summarize all our findings in section \ref{sec:conclusion}.

\vspace{-0.3cm}

\section{U(1) Lattice Gauge Theory}

\subsection{Gauge Action}
\label{sec:lgt}
\vspace{-0.2cm}
We consider a  hypercubic lattice with spacetime coordinates labelled as $n=(\tau,r)$, where $r$ labels points on the spatial lattice, and $\tau$ the time slice. Directions are labelled as $\mu, \nu = (0,1,2,3)$ with $\mu = 0$ for the temporal direction. 
The canonical $\mathrm{U}(1)$ lattice gauge theory action may be written in terms of angles $\phi_\nu (n)\in [0,2\pi)$, parameterising $\mathrm{U}(1)$ group elements $U_\nu(n) = {\rm e}^{{\rm i} \phi_\nu (n) }$ on links of the lattice. The lattice action is
\begin{equation}
S[\phi]=-\beta \sum_{n, \mu<\nu} \cos \Theta_{\mu\nu}(n)  \, ,
\label{eq:lgt:actionderivedhamiltonian1}
\end{equation}
where $\Theta_{\mu\nu}(n) = {\phi}_{\mu}(n)+{\phi}_{\nu}(n +\hat{\mu}) - {\phi}_{\mu}(n  +\hat{\nu}) -{\phi}_{\nu}( n)$ are terms living on plaquettes. The corresponding quantum theory is defined through the generating function $Z = \int D\phi\,  e^{-S[\phi]}$. The theory has a $U(1)$ gauge invariance under $U_\nu(n)\rightarrow \eta(n) U_\nu(n) \eta^{\dagger}(n+\nu)$ where $\eta(n)$ are local phases. 

We define the theory on a lattice with $N_s$ sites along each of the three spatial dimensions, and $N_t$ sites along the temporal direction. Periodic boundary conditions are imposed in all directions. 

In the following, it will be convenient to generalise the action in Eq.\eqref{eq:lgt:actionderivedhamiltonian1} to anisotropic lattices with different spatial $a_s$, and temporal, $a_t$ lattice constants. 
Defining the anisotropy parameter  $\xi = a_s/a_t$, spatial and temporal lattice directions have to be treated differently at the level of the action, yielding \cite{ENGELS1982545},
 \begin{equation}
S[\phi]=-\frac{\beta}{\xi} \sum_{n, l<l'} \cos \Theta_{ll'}(n) 
- \beta \xi \sum_{n, l}\cos\Theta_{l0}(n) \, ,
\label{eq:lgt:actionderivedaniisotropic1}
\end{equation}
with $l,l'=1,2,3$ denoting the spatial directions. 

\vspace{-0.3cm}
\subsection{Observables}
\vspace{-0.2cm}

In a pure U(1) LGT all observables, i.e. gauge-invariant quantities, can be constructed as the trace of the product of link variables across closed loops \cite{Gattringer:2010zz}.
Of particular interest for this work is the Polyakov loop at the site $r$, 
\begin{equation}
    P(r)=\prod_{j=0}^{N_{T}-1} U_{0}(r, j) \, ,
\end{equation}
obtained as the product of temporal link variables along the full extent of the temporal lattice. It corresponds to a  closed loop for the considered periodic boundary conditions. 
The average Polyakov loop,
\begin{equation}
    P = \frac{1}{N_s^3}\sum_r P({r}) \, ,
\end{equation}
can be interpreted as the probability of observing a single static charge \cite{Gattringer:2010zz} over the probability that no charge is observed. 

Therefore, this observable distinguishes confined and deconfined phases. At large values of the coupling constant (small $\beta$) we expect the theory to be confined, so that no free charges are observed, i.e. $P=0$ at the thermodynamic limit. At small values of the coupling constant (large $\beta$), we expect the theory to be deconfined, so that free static charges can be observed, i.e. $P \neq 0$. Therefore, the average Polyakov loop can be taken as the order parameter for the confined/deconfined transition. In the following, we use $P$ to determine the value critical value $\beta_c$ at which the phase transition occurs.

\begin{figure}[h]
    \centering
    \vspace{-0.7cm}
    \includegraphics[width=0.95\columnwidth]{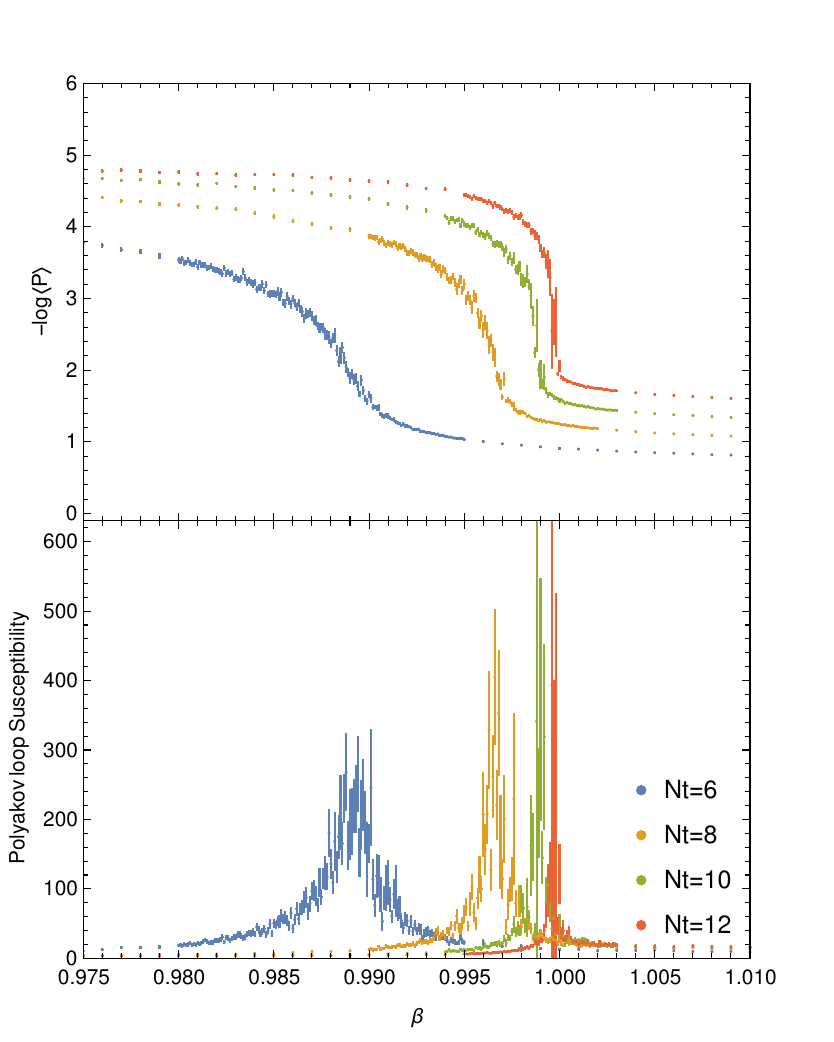}
    \vspace{-0.4cm}
    \caption{Logarithm of the average value of the Polyakov loop (top) and Polyakov loop susceptibility (bottom) as a function of $\beta$ for $N_t = 6, 8, 10, 12$ and $N_s = 24$, for an anisotropic lattice with $\xi = 1.5$. The average Polyakov loop is $P = 0$ in the confining phase, and changes to a non-zero value in the Coulomb phase. The peak in the susceptibility identifies the location of the phase transition.}
    \label{fig:ch:results:polyakov1}
\end{figure}

For a finite size lattice this translates to the behavior illustrated in the upper panel of Fig.~\ref{fig:ch:results:polyakov1}. For large values of $\beta$ we observe $\log \left\langle P \right\rangle  \propto - N_t$ whereas for small $\beta$ the asymptotic value reaches an $N_t$ independent constant.

In order to better identify the phase transition, we study the susceptibility of the Polyakov loop, defined as 
\begin{equation}
    \chi_P = N_t N_s^3(\langle P^2\rangle - \langle P\rangle^2) \, ,
\end{equation}
which peaks at the critical value of $\beta$ for which the phase transition occurs, as shown in the lower panel of Fig. \ref{fig:ch:results:polyakov1}.

\vspace{-0.3cm}
\subsection{Simulating the U(1) LGT}
\label{sec:results}
\vspace{-0.2cm}

To determine the average value of the Polyakov loop and its susceptibility, we carried out GPU accelerated Metropolis-Hastings Monte Carlo \cite{Metropolis1953, Hastings:1970aa, Cardoso:2010di, Cardoso:2011xu, Amado:2012wt, Bicudo:2021tsc}. We generated $N_{\rm iter} = 10^5$ Markov Chain iterations and dropped the first 5000, in order to consider only iterations generated after the thermalization of the system. We further calculated the autocorrelation time, $\tau$, between the configurations, and kept only configurations separated by $3\tau$. In order to decrease the autocorrelation between two sequential configurations, three over-relaxation steps were implemented \cite{Creutz:1987xi}. Autocorrelation times between $\tau = 1.4$ and $\tau = 15$ were obtained in different regions of the parameter space, resulting in a final number of configurations between $N_{\rm config} = 22.6 \times 10^4$ and $N_{\rm config} = 2.1 \times 10^3$. The variance of the observables was calculated with the Jackknife method \cite{10.1093/biomet/43.3-4.353,10026575637}.

From these simulations taken for different values of $N_t$ and $\xi$, we obtain the critical value of $\beta(N_t,\xi)$ for which the confinement/deconfinement phase transition occurs by performing a Lorentzian fit to the Polyakov loop susceptibility.
We have checked that results do not change significantly with $N_s$ for $N_s$ sufficiently large.

\subsection{ Effective Hamiltonian }
\vspace{-0.2cm}

To link the coupling constants of the lattice gauge theory defined in the previous section to physical parameters we shall assume that there is a  Hamiltonian for the gauge theory with couplings $U$ and $K$ that are to be determined from the lattice action results \cite{Kogut1975}, 
\begin{equation}
H =\frac{U}{2} \sum_{r, l}\left({n}_{l}(r)\right)^{2}-K \sum_{r, l< l'} \cos \left[\Theta_{ll'}(r)\right] \, ,
\label{hamiltonian_u1}
\end{equation}
where ${n}_{l}(r)$ an integer-valued operator conjugated to ${\phi}_{l}(r)$. 

\subsection{ Parameter Matching }
\vspace{-0.2cm}

Starting from the Hamiltonian of Eq.\eqref{hamiltonian_u1}, the standard derivation of the partition function $Z=\text{Tr}[e^{-H/T}]$ by Trotter slicing, that we sketch in appendix \ref{appa} for completeness, relies on the Villain approximation to integrate the conjugated variables, ${n}_{\mu}(r)$. 
A naive identification of the resulting action, valid for large values of  $T N_t / U $, and that of Eq.\eqref{eq:lgt:actionderivedaniisotropic1} yields 
\begin{eqnarray}  
\frac{K}{T N_t} = \frac{\beta}{\xi}, \label{relations_k_tnt_initial}\\ 
\frac{T N_t}{U} = \beta \xi.
\label{relations_t_u_initia}
\end{eqnarray}
However, at small $N_t$, there are important corrections at low temperatures.
A first refinement to Eqs. \ref{relations_k_tnt_initial} and \ref{relations_t_u_initia} can be obtained using a perturbative expansion \cite{Janke1986}, which we provide in Appendix \ref{appb}. As we show in Appendix \ref{appb}, despite clear improvements over the naive parametrization Eqs.~\ref{relations_k_tnt_initial} and \ref{relations_t_u_initia}, this approach is still not sufficient to determine the phase diagram given the system sizes available to us.

Nevertheless, an action of the form of Eq.\eqref{eq:lgt:actionderivedaniisotropic1} is expected to capture the physics of $H$, especially in the vicinity of the phase transition where irrelevant terms can be neglected.   

With this motivation, we renormalize the parameters in the action, replacing the coefficient multiplying the
space-time part with a scaling function of $T N_t / U$ to be determined, i.e.  
\begin{equation}
S[\phi]=-  \frac{K}{T N_t} \sum_{n, \mu<\nu} \cos \Theta_{\mu\nu}
- f\left( \frac{T N_t}{U} \right) \sum_{n, \mu}\cos\Theta_{\mu0}  \,.
\label{eq:lgt:actionderivedhamiltonian1}
\end{equation}

Identifying the terms with those in Eq.\eqref{eq:lgt:actionderivedaniisotropic1} and solving in order to $K/U$ and $T/U$ we find 
\begin{eqnarray}  
\frac{K}{U} = \frac{\beta   }{\xi } f^{-1}\left(\beta\xi\right),
\label{eq:KoU_1} \\
\frac{T}{U} = \frac{1}{ N_t} f^{-1}\left(\beta\xi\right) \, .
\label{eq:ToU_1}
\end{eqnarray}
To determine $f(z)$ numerically we first approximate the quantity $\beta_0(\xi) = \lim_{N_t\to\infty} \beta(N_t,\xi)$ by $\beta(N_t,\xi)$ computed with the largest available value of $N_t$ and confirm that appropriate convergence with $N_t$ was achieved. 

\begin{figure}[t]
    \centering
    \includegraphics[width=0.9\columnwidth]{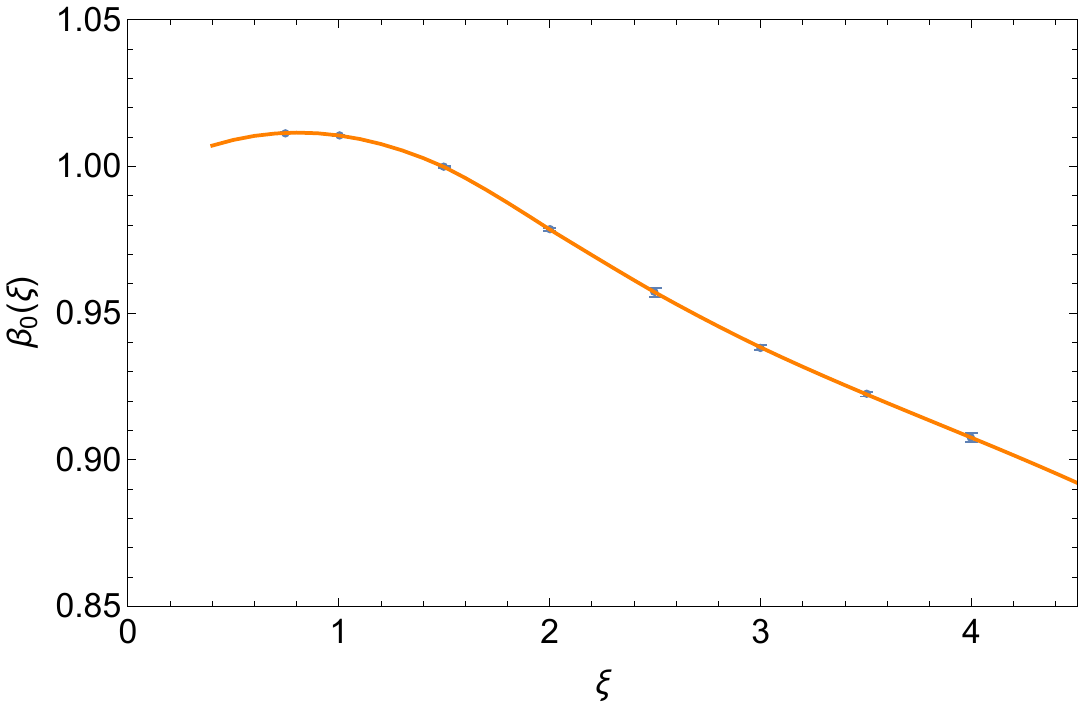}
    \caption{Values of $\beta_0(\xi) = \lim_{N_t\to\infty} \beta(N_t,\xi)$ for each value of $\xi$ approximated using the largest value of $N_t$ available.}
    \label{fig:ch:results:beta0_xi}
\end{figure}

\begin{figure}[t]
    \centering
    \includegraphics[width=0.9\columnwidth]{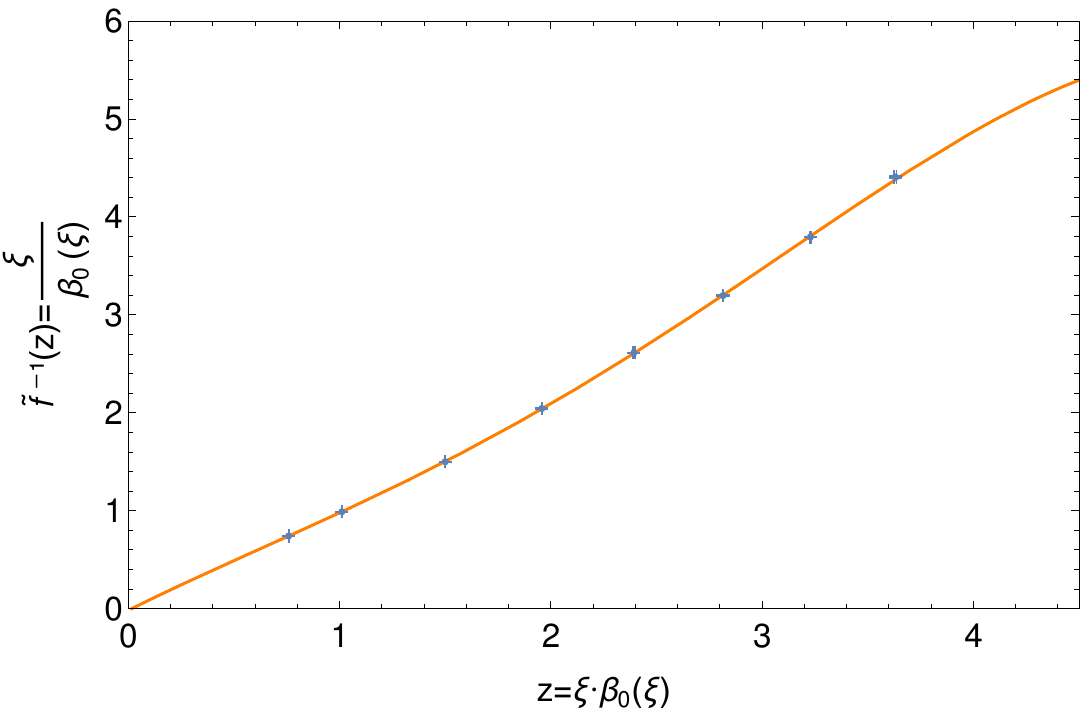}
    \caption{Function $\tilde{f}^{-1}(z)$ obtained by fitting a curve to $z = \xi \beta_0\left(\xi\right)$ and $\tilde f ^{-1} = \xi/\beta_0(\xi)$, according to Eq. \ref{eq:results:K_0oU} and using the values of $\beta_0$ determined for each value of $\xi$, as shown in Fig \ref{fig:ch:results:beta0_xi}}
    \label{fig:ch:results:f_beta0_xi}
\end{figure}

The critical value of $K=K_0$ at $T=0$ thus given by Eq.\eqref{eq:KoU_1} evaluated at $\beta_0(\xi)$
\begin{equation}
    \frac{K_0}{U} = \frac{\beta_0(\xi)   }{\xi }f^{-1}\left[\beta_0(\xi)\xi\right]. 
    \label{eq:results:K_0oU}
\end{equation}
Since the left hand side of this equality is independent of $\xi$ and since $\beta_0(\xi)$ was obtained previously,  this relation can be used to determine $f^{-1}(z)$ up to a multiplicative constant, i.e. the function $\tilde f ^{-1}(z) = U f^{-1}(z)/K_0$.  
The quantities $\beta_0(\xi)$ and $\tilde f ^{-1}(z)$ are shown in Figs. \ref{fig:ch:results:beta0_xi} and \ref{fig:ch:results:f_beta0_xi}, respectively.

Given $\tilde f$, the mapping between the action and the Hamiltonian is completely determined by 
\begin{eqnarray}  
\frac{K}{K_0} = \frac{\beta  }{\xi }  \tilde f^{-1}\left(\beta\xi\right),
\label{eq:KoU} \\
\frac{T}{K_0} = \frac{1}{ N_t} \tilde f^{-1}\left(\beta\xi\right).
\label{eq:ToU}
\end{eqnarray}
A drawback of this procedure is that it only determines the mapping between  Hamiltonian and action coupling constant up to a constant, $K_0$, that has to be determined independently.

\begin{figure}[h]
    \centering
    \includegraphics[width=0.95\columnwidth]{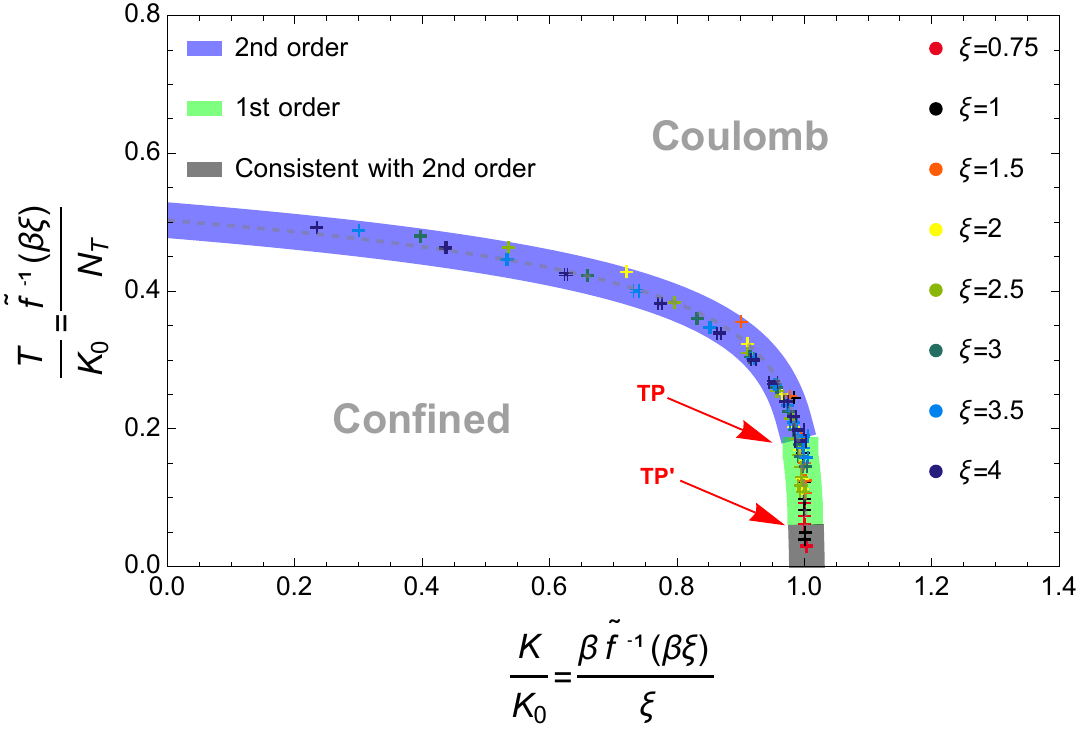}
    \caption{Phase diagram using the parameters in Eqs. \ref{eq:KoU} and \ref{eq:ToU}. Different markers represent different extents of the temporal lattice, $N_t$. The error bars were computed and are smaller than the markers. The calculated values of $T/K_0$ and $K/K_0$ for each value of $\xi$ and $N_t$ are presented in appendix \ref{app_table}, in table \ref{appa:tab1}. TP indicates the tricritical point at which the transition changes between first and second order. TP' instead refers to the point below which the transition is consistent with second order.}
    \label{fig:results:pdfx}
\end{figure}

\vspace{-0.3cm}

\section{Results}

\subsection{Phase diagram}
\label{sec:pd}
\vspace{-0.2cm}

Fig. \ref{fig:results:pdfx} shows the phase diagram of the U(1) gauge Hamiltonian \eqref{hamiltonian_u1}, obtained via the approach outlined in the previous section. Using our parameter matching procedure, the critical values of $\beta(N_t,\xi)$, obtained by simulating the lattice action for many different $(N_t,\xi)$ couplings, collapse into a single curve.  

The quality of the data collapse is remarkable.
As expected, we find that for large $T$ the transition is second order and passes to first order at a tricritical point, TP, as $T$ decreases. 
Interestingly, for even lower temperature, we find that the discontinuity becomes weaker. In fact, our data is compatible with a second order transition for $T=0$. 
These findings are substantiated in the next section.

\subsection{Order of the Phase Transition}
\label{sec:opt}
\vspace{-0.2cm}

We now turn to the discussion of the order of the phase transition along the entire phase boundary. 
To identify the order of the transition, we examine the histograms of the absolute value of the Polyakov loop on both sides of the transition.
The first order nature of the transition is determined from phase co-existence signalled by a double peak structure in the histogram \cite{Cardoso:2011hh}. 

Histograms are obtain using the field configurations of $N_{\rm iter} = 200000$ consecutive Markov Chain iterations and discarding the first $5000$ in order to remove the pre-thermalization regime.
Simulations were carried out with spatial extents $N_s = 24$, and $N_s = 36$. 

\begin{figure}[h]
    \centering
    \includegraphics[width=0.95\columnwidth]{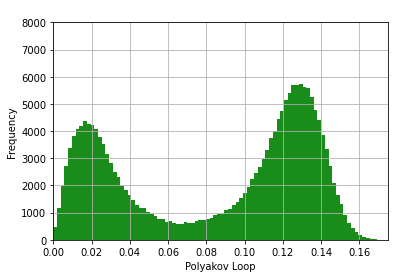}
    \caption{Histogram of the Polyakov loop at the phase transition for $T/K_0 = 0.12$. The two peaks identify phase coexistence at a first order transition.}
    \label{fig:results:hist_example_1st}
\end{figure}

\begin{figure}[t]
    \vspace{-0.4cm}
    \centering
    \includegraphics[width=0.9\columnwidth]{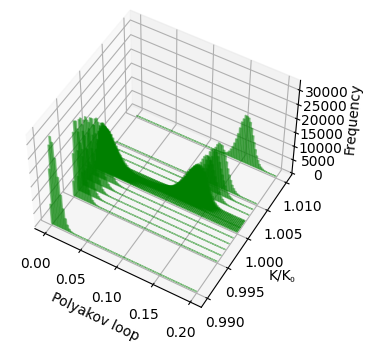}
    \caption{Histograms of the Polyakov loop for $T/K_0 = 0.12$ and varying $K/K_0$. Close to the phase transition, we can identify two peaks in the histogram.}
    \label{fig:results:hist3d_example_1st}
\end{figure}

Fig.~\ref{fig:results:hist_example_1st} illustrates the two peak structure clearly observed at the phase transition for $T/K_0 = 0.12$. 
Fig.~\ref{fig:results:hist3d_example_1st} shows how the Polyakov loop histogram evolves across the phase transition from the confined to the decofined phases, for fixed $T/K_0$ and various $K/K_0$. In the confined phase the histogram is single-peaked. The two peak structure appears around the phase transition and vanishes again deeper inside the deconfined phase.

We identify a two-peak structure for temperatures $0.05 < T/K_0 \leq {0.175}$ consistent with a discontinuous change of the Polyakov loop as the system transitions from the confined to the Coulomb phase. Accordingly we label transitions in this region as first order.

\begin{figure}[t]
    \vspace{-0.4cm}
    \centering
    \includegraphics[width=0.9\columnwidth]{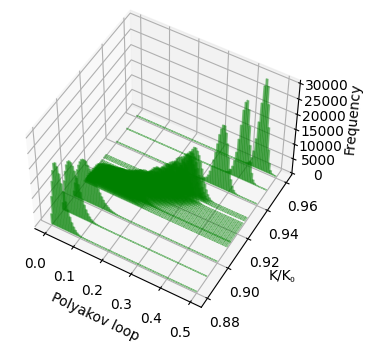}
    \caption{Histograms of the Polyakov loop at $T/K_0 = 0.32$ and varying $K/K_0$.}
    \label{fig:results:hist3d_example_2nd}
\end{figure}

Fig.~\ref{fig:results:hist3d_example_2nd} shows the evolution of the histogram across the phase transition for a temperature $T/K_0 \geq {0.19}$. Here, only one peak can be identified in the histogram that is progressively centred at higher values of the Polyakov loop. This continuous change of the average values of the Polyakov loop is compatible with a second order phase transition. 

From the analysis of the histogram across the transition, our data is consistent is the existence of a temperature point between $T/K_0 = 0.175$ and $T/K_0 = 0.19$ at which the transition changes from first to second order. This high temperature change in the order of the phase transition had already been explored using isotropic lattices as we now briefly summarize. 

In \cite{Vettorazzo2004_2}, the authors analysed time histories of the plaquette mean values and found signs of metastability on isotropic lattices with $N_t = 6$ and $N_t = 8$, indicating that the transition is of first order for lattices with these extents in the temporal direction. For a smaller lattice (higher temperature) their results seem to indicate that the transition could be second order for $N_t = 4$.
More recently, in \cite{PhysRevD.88.065025}, the authors conclude that the transition is first order for lattices with $N_t \geq 6$ and second order for lattices with $N_t \leq 5$ by studying the scaling of the plaquette and Polyakov loop susceptibilities with $N_s$, using isotropic lattices with fixed values of $N_t$. 

The results we report above are in accordance with these previous studies. Moreover, by considering the anisotropic lattice regularization, we are able to explore a wider parameter region, and obtain a comprehensive coupling-temperature phase diagram that reveals the tricritical point connecting the highest temperature regime ($T/K_0>T_{\text{TP}}$) to intermediate temperatures ($0.05<T/K_0<T_{\text{TP}}$). Our numerical analysis of the double-peak structure places the tricritical temperature, $T_{\text{TP}}$, within the interval $0.175<T_{\text{TP}}<0.19$. In addition, our method is also able to access the previously unexplored lower temperature regime that we discuss below. 

\begin{figure}[t]
    \centering
    \includegraphics[width=0.95\columnwidth]{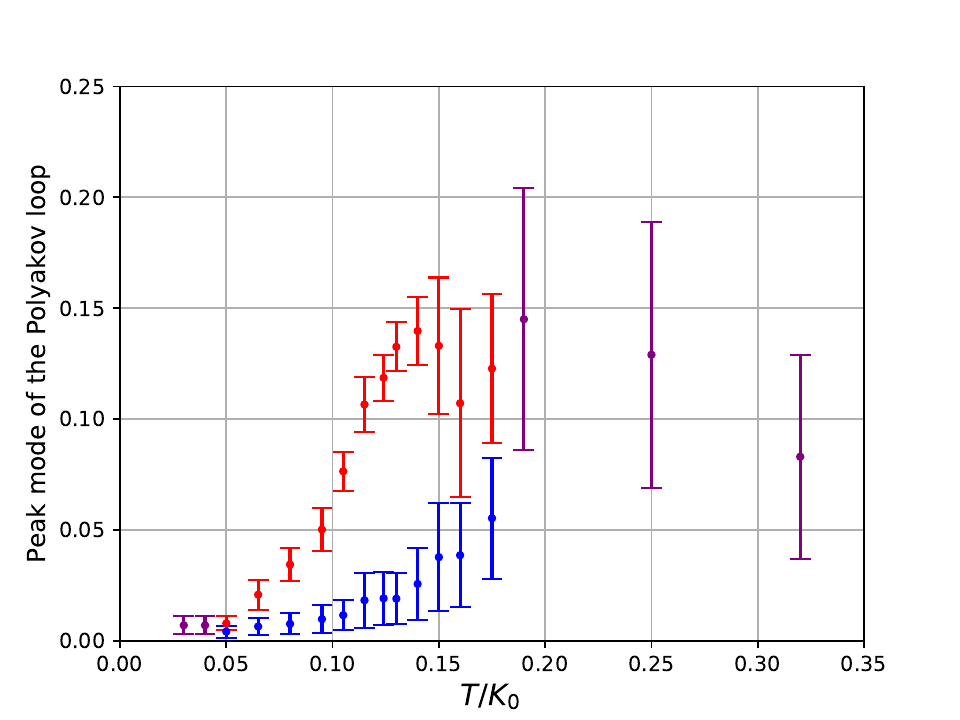}
    \caption{Position of the peaks in the Polyakov loop histogram for the values of $T/K_0$ considered. For the cases where two peaks were identified, both peak positions are plotted in blue and red. For the cases where only one peak was identified, the position of the peak at the phase transition is plotted in purple.}
    \label{fig:results:peak_position}
\end{figure}

\begin{figure}[t]
    \centering
    \includegraphics[width=0.95\columnwidth]{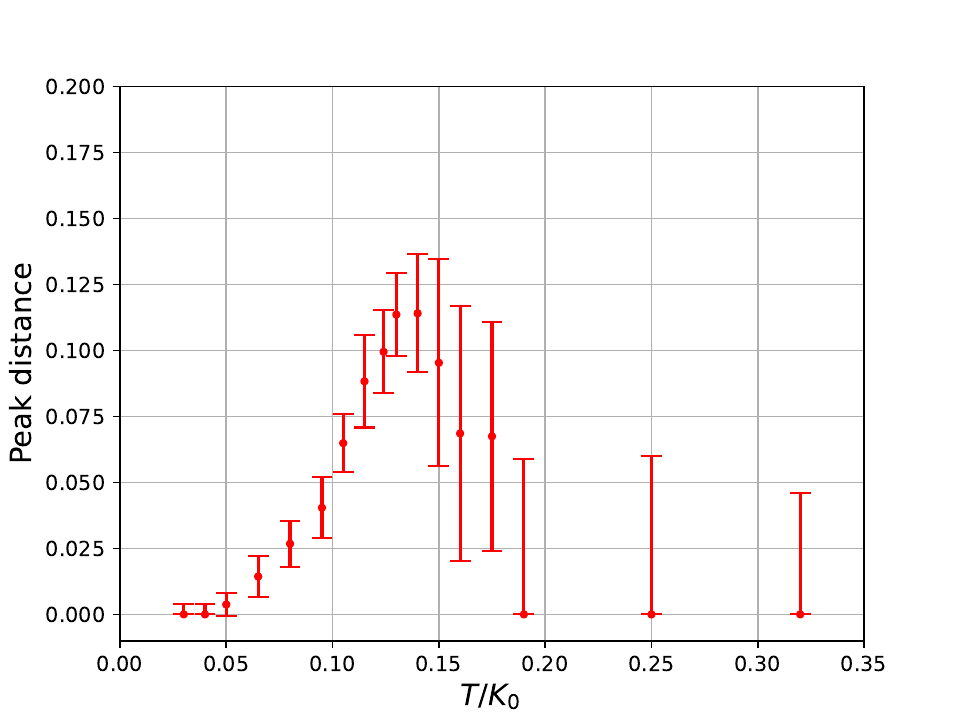}
    \caption{Distance between the peaks in the Polyakov loop histogram. In the cases where only one peak was identified, the peak distance is plotted at zero and the error bars correspond to the error in the peak position.}
    \label{fig:results:peak_distance}
\end{figure}

Fig.~\ref{fig:results:peak_position} shows the position of the peaks in the Polyakov loop histogram versus temperature for various values of $T/K_0$. The distance between peaks is plotted in Fig.~\ref{fig:results:peak_distance}. As expected, from moderate to high temperatures, the peak-to-peak distance vanishes (up to error bars) concomitantly with transition change from first to second order. 
Surprisingly, for low temperatures, the distance between the peaks also vanishes as the temperature decreases. Indeed, for $T/K_0 \leq 0.05$, the peak-to-peak distances vanish within error bars. Although it is impossible to exclude a very weak first order transition, these results strongly indicate the existence of a low-temperature tricritical point, $\text{TP'}$, where the transition apparently becomes second order from $T_\text{TP'}/K_0 \simeq 0.05 $ down to $T=0$.

As previously mentioned, there has been some debate about the order of the phase transition of the U(1) LGT in the zero temperature limit. This question has been addressed using isotropic lattices \cite{Jersak1983, Evertz1985, Barber1984,Arnold2003, Vettorazzo2004} with an apparent consensus that the transition is first order in the limit of zero temperature. 
However, isotropic lattice regularization is constrained to specific values of temperature, whereas anisotropic lattices allow one to explore a much wider range of temperatures. In particular, the lattice sizes used in these previous studies \cite{Jersak1983, Evertz1985, Barber1984,Arnold2003, Vettorazzo2004} correspond to temperatures greater than $T/K_0 = 0.05$, for which our method also predicts a first order transition. 
It is only for values of the temperature below those considered in previous studies that we find the possibility that the two peaks in the Polyakov loop histogram merge implying a continuous transition below $TP'$ and down to zero temperature.

In Fig.~\ref{fig:results:pdfx}, we present the complete phase diagram with the order of the transition indicated.

\subsection{Critical exponents}
\label{sec:exp}
\vspace{-0.2cm}

In order to examine the nature of the phase transition in the regions $T/K_0 \geq 0.19$ and $T/K_0 \leq 0.05$, we investigated the critical exponents for points in each of these regions. 

We calculated the Binder cumulant, 
\begin{equation}
    B = 1 - \frac{{\langle P ^4\rangle}}{3{\langle P ^2 \rangle}^2} \, 
\end{equation}
and the Polyakov loop susceptibility, $\chi_P$, for fixed values of $T/K_0$ and varying $K/K_0$, for several values of the spatial volume, while keeping the temporal size, $N_t$, fixed. Near the phase transition, these quantities scale as
\begin{equation}
\begin{aligned}
    \chi_P &\sim  {N_s}^\frac{\gamma}{\nu} f(\Delta K \cdot {N_s}^\frac{1}{\nu} )\\
    B &\sim  f(\Delta K \cdot {N_s}^\frac{1}{\nu} ) \, ,
\end{aligned}
\end{equation}
where $\Delta K = (K - \left.K\right\rvert_{\textrm{crit}} )/K_0 $,  with $(K/K_0)_{\textrm{crit}}$  the critical value of $K/K_0$ at infinite spatial volume, and $\gamma$ and $\nu$ are the critical exponents.

\begin{figure}[t]
    \vspace{-0.5cm}
    \centering
    \includegraphics[width=\columnwidth]{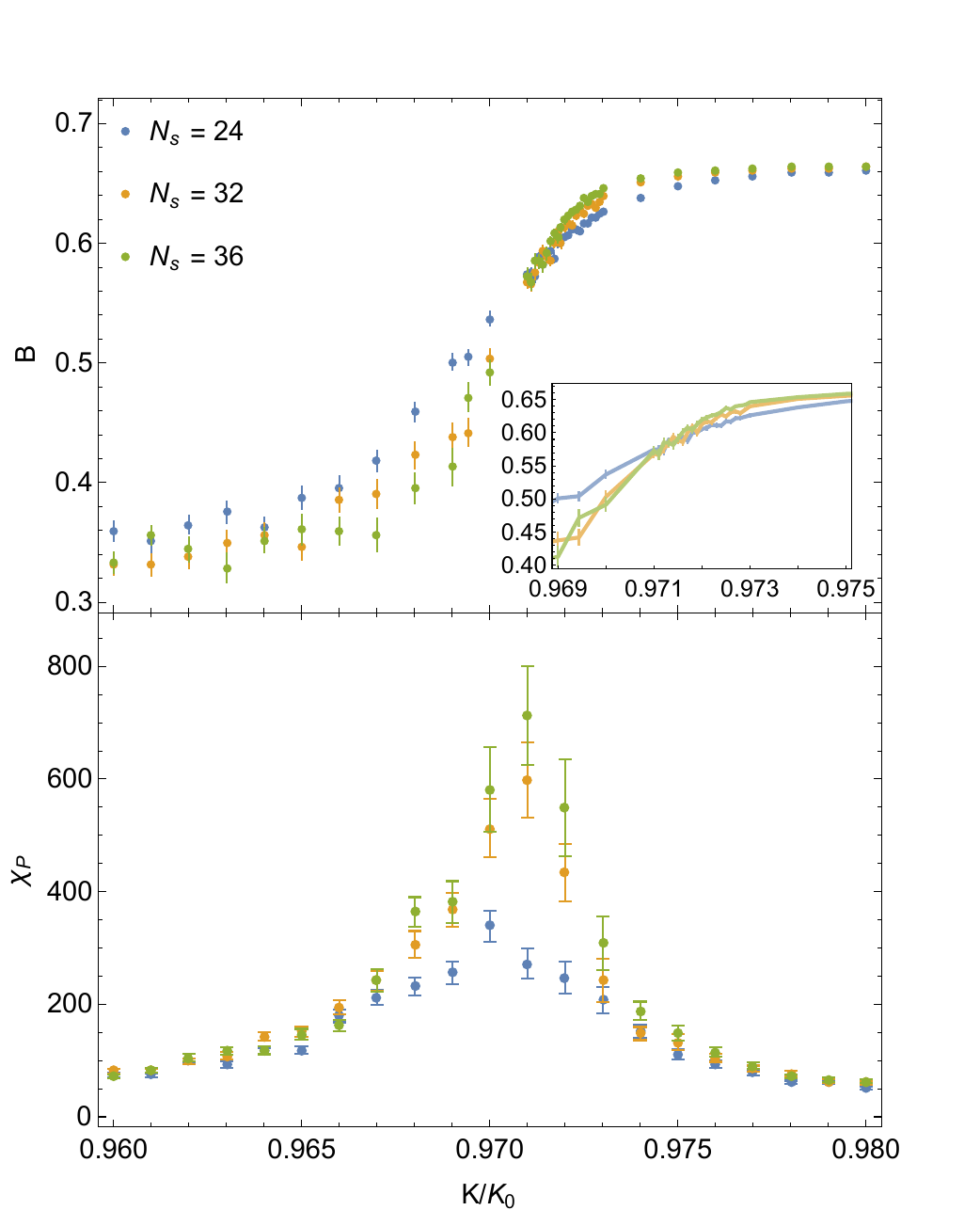}
    \caption{Binder cumulant (top) and Polyakov loop susceptibility (bottom) at $T/K_0 = 0.25$ and varying $K/K_0$, for $N_t = 8$ and several values of $N_s$. The curves of the Binder cumulant for different $N_s$ cross at $(K/K_0)_{\textrm{crit}} = 0.9715 \, \pm \, 0.0005$, indicating the critical value of $K/K_0$ at infinite spatial volume.}
    \label{exp:n2_binder_polysus}
\end{figure}

\begin{figure}[t]
    \vspace{-0.5cm}
    \centering
    \includegraphics[width=\columnwidth]{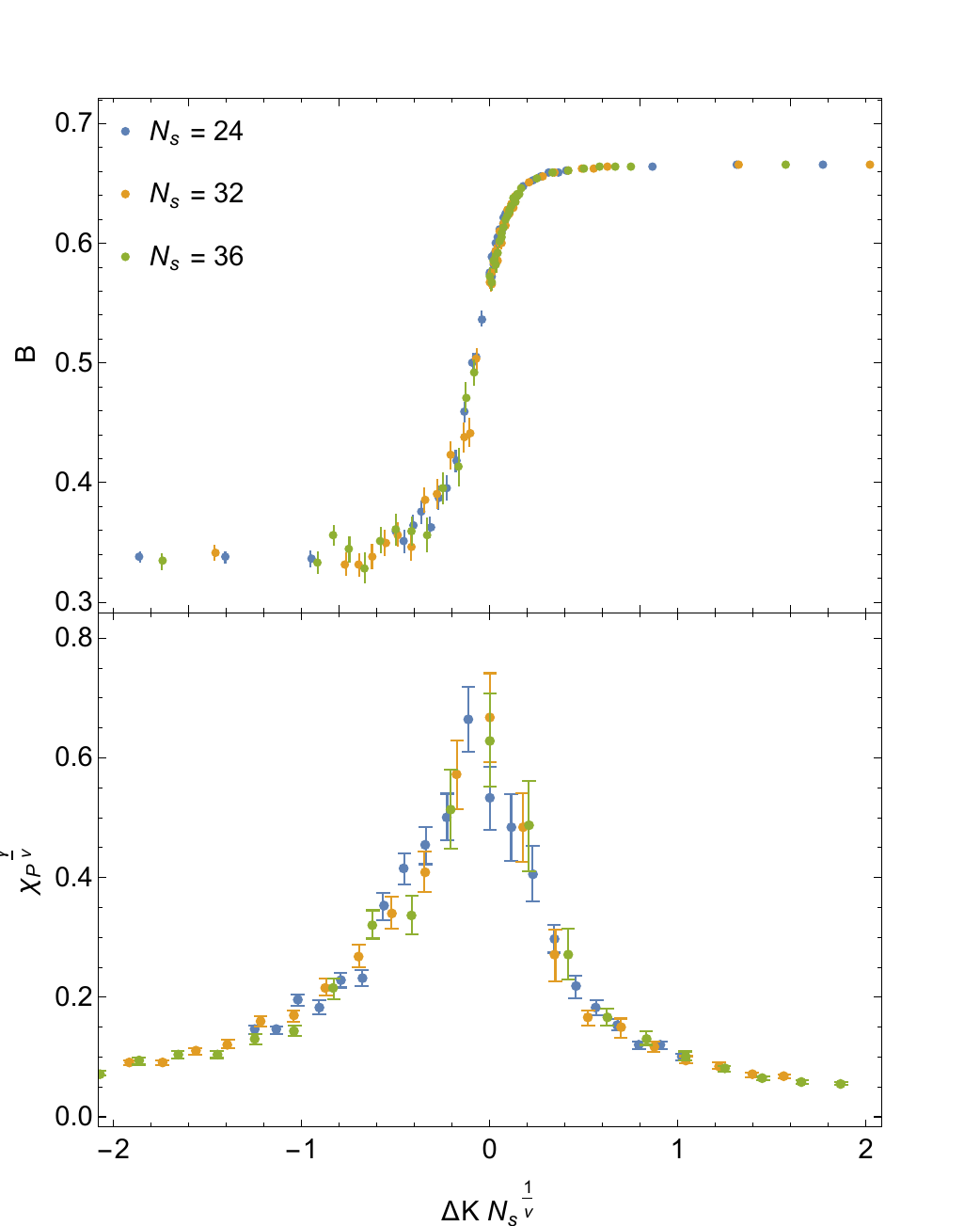}
    \caption{Data collapse of the Binder cumulant (top) and the Polyakov loop susceptibility (bottom) at $T/K_0 = 0.25$ for different spatial sizes obtained for $\nu = 0.67171$ and with $(K/K_0)_{\textrm{crit}} = 0.9715$.}
    \label{exp:n2_binder_polysus_coll}
\end{figure}

Fig.~\ref{exp:n2_binder_polysus} shows the Polyakov loop susceptibility and the Binder cumulant obtained at $T/K_0 = 0.25$. We estimated the critical value $(K/K_0)_{\textrm{crit}}$ from the crossing of $B$ for different system sizes.  Fig.~\ref{exp:n2_binder_polysus_coll} depicts the same data scaled with the critical exponents for the 3D XY universality class, $\nu = 0.67171$ and $\gamma = 0.13178$ \cite{PhysRevD.88.065025}. The data collapse indicates that the high temperature phase transition of the $U(1)$ lattice gauge theory is of the 3D XY universality class as one would expect as the model maps to the 3D XY model in the $K/K_0 = 0$ limit.

\begin{figure}[t]
    \centering
    \includegraphics[width=0.95\columnwidth]{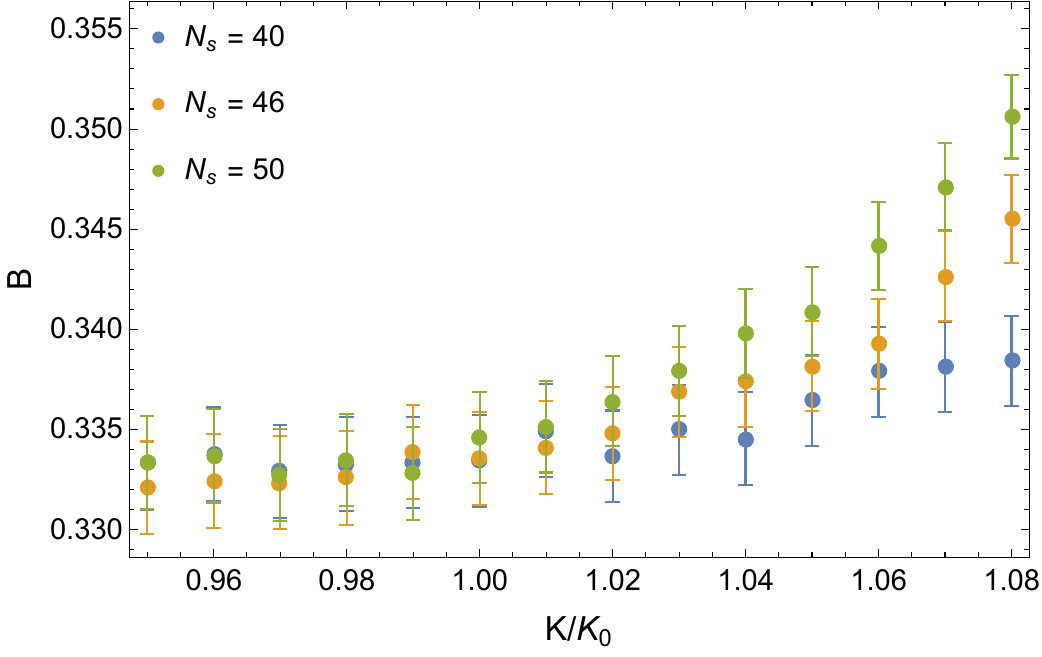}
    \caption{Binder cumulant at $T/K_0 = 0.03$ for different spatial sizes. No discontinuity is observed at the phase transition, which is consistent with a second order transition.}
    \label{exp:t003_binder}
\end{figure}

Motivated by the fact that the peak distance in the Polyakov loop histogram goes to zero, we examine the Binder cumulant in order to address the possibility of a second order transition at low temperatures ($T/K_0 < 0.05 $). The transition is harder to probe at low temperatures  as the onset of the Binder cumulant is more shallow. We studied $T/K_0 = 0.03$ where the discontinuity in the Binder cumulant is negligible and where the histogram reveals a single peak. Taking  large values of the spatial lattice extent ($N_s = 36,40,46$), and with $N_{\rm iter} = 1.6 \times 10^7 $ we obtain Fig.~\ref{exp:t003_binder} that shows the Binder cumulant varying smoothly across the phase transition which is consistent with a second order phase transition.

\begin{figure}[t]
    \centering
    \includegraphics[width=0.95\columnwidth]{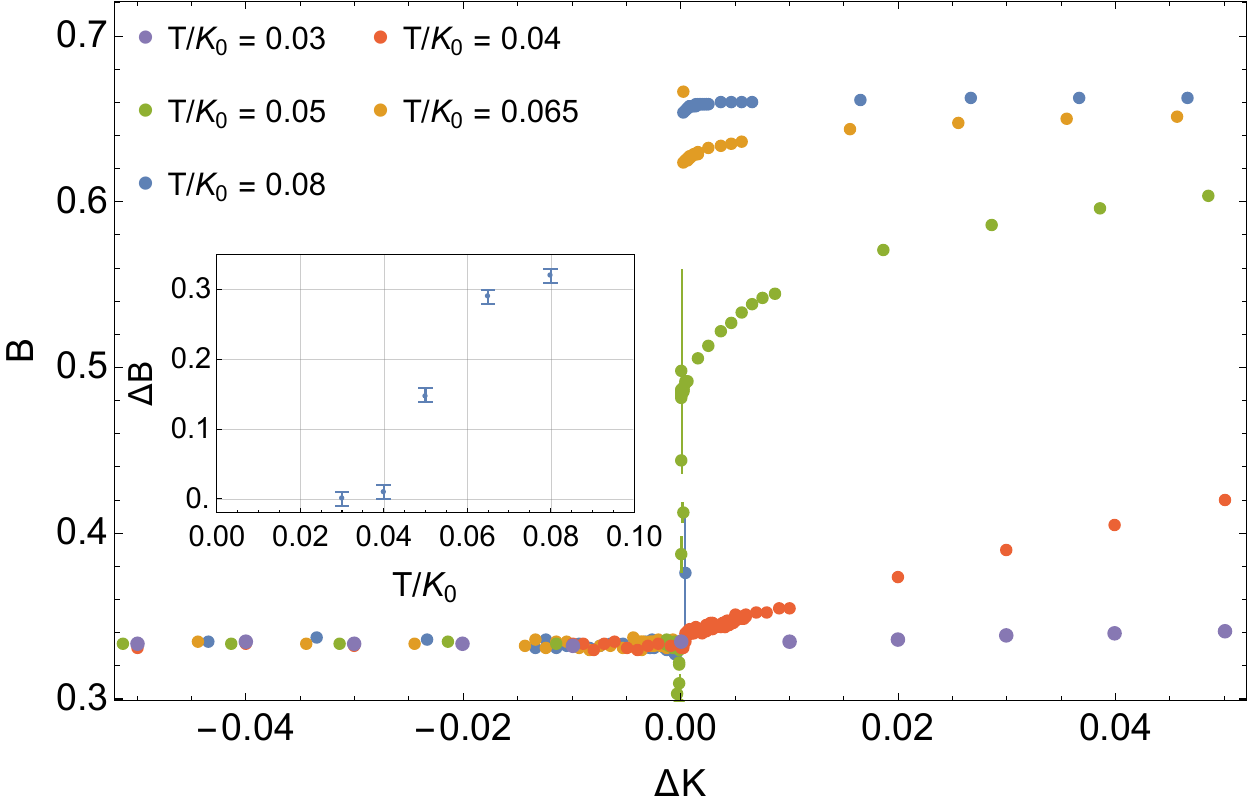}
    \caption{Binder cumulant for several values of $T/K_0$ calculated at the highest value of $N_S$ considered for each temperature. The discontinuity decreases with decreasing temperature, indicating that the transition becomes weakly first order.}
    \label{exp:discontinuity2}
\end{figure}

The collapse of the Binder cumulant data strongly depends on the value of $(K/K_0)_{\textrm{crit}}$ found. With the data shown in Fig.~\ref{exp:t003_binder}, it is difficult to specify precisely the point at which the curves for different spatial system sizes cross, and values of $(K/K_0)_{\textrm{crit}} \in [ 0.96, 1 ] $ are compatible with the data obtained. Motivated by the values of $(K/K_0)_{\textrm{crit}}$ obtained for higher temperatures, we take $(K/K_0)_{\textrm{crit}} = 0.97$ for $T/K_0 = 0.03$, and explore the values of the critical exponent $\nu$ that allow for the data to be collapsed.  

\begin{figure}[h]
    \centering
    \vspace{0.5cm}
    \includegraphics[width=0.95\columnwidth]{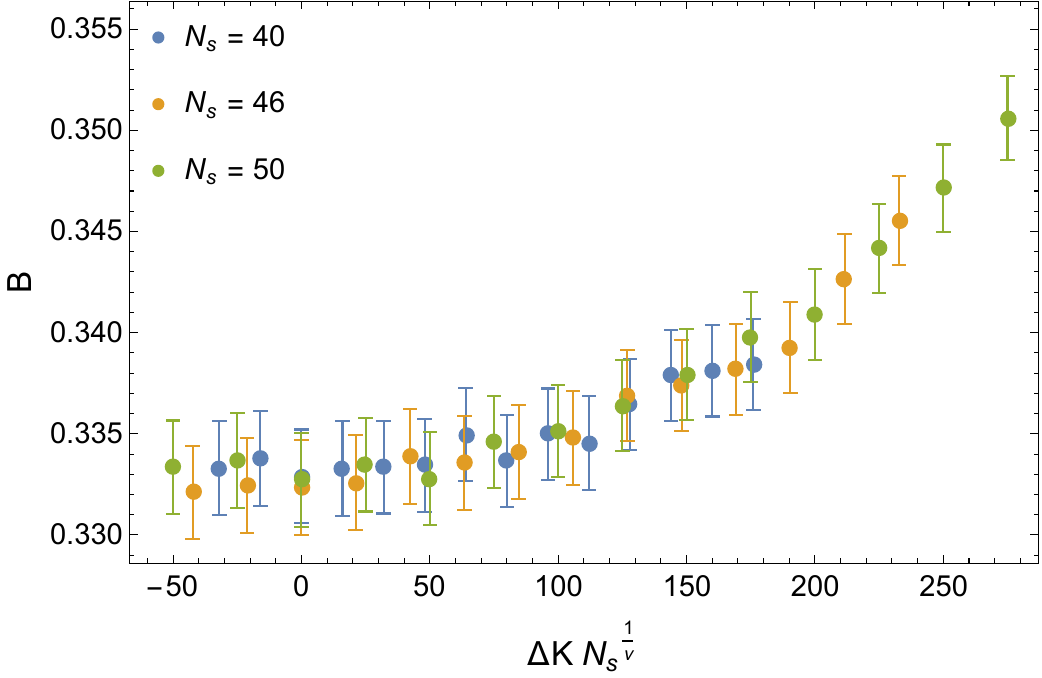}
    \caption{Collapse of the Binder cumulant data at $T/K_0 = 0.03$ obtained for $\nu = 0.5$ and $(K/K_0)_{\textrm{crit}} = 0.97$.}
    \label{exp:t003_binder_coll}
\end{figure}

To illustrate this, let us suppose that the transition is in the XY universality class. In four dimensions, the XY model is at its upper critical dimension so, for this scenario, we would expect mean field exponents with logarithmic corrections. Fig.~\ref{exp:t003_binder_coll} shows an instance of a collapse for mean field exponent $\nu=1/2$. Similar such plots are obtained for $\nu$ in the range $0.3-0.7$   illustrating that the Monte Carlo data are consistent with a continuous transition below a temperature of $T/K_0 = 0.05$ with $\nu$ in the range $0.3-0.7$ that includes the mean field exponent.

\vspace{-0.3cm}
\section{Conclusion}
\label{sec:conclusion}

In this paper we have revisited the phase diagram of the $4D$ $\mathrm{U}(1)$ LGT using GPU accelerated Monte Carlo simulations on anisotropic lattices. Previous simulations had hinted at the presence of a tricritical point along the phase boundary separating the confined and deconfined phases.

We explored a range of lattice anisotropies with a rescaling of the couplings in the action to obtain phase diagram parameterized by the Hamiltonian couplings. This allows one to explore the phase diagram as a function of coupling $\beta$ and temperature with considerable resolution in both couplings. We determined that there is a first order region at $0.05 < T/K_0 \leq 0.175$, and a  second order region for $T/K_0 \geq 0.19$ with critical exponents consistent with those of the 3D XY universality class. These simulations clearly identify a tricritical point at intermediate temperatures between $T/K_0 = 0.175$ and $T/K_0 = 0.19$. For the region $T/K_0 \leq 0.05$ we find that the transition becomes more and more weakly first order as the temperature is lowered. Our simulations are compatible both with a continuous transition and a very weakly first order transition in the zero temperature limit. 

This paper reveals rich structure of the phase diagram in the pure gauge theory in four dimensions. If one switches on a coupling to scalar matter in the fundamental representation \cite{Fradkin1978} the deconfined phase extends to finite matter coupling terminating in a tricritical point at the intersection of a first order line, the Higgs transition and the confinement/deconfinement transition. It is of interest to explore the evolution of this phase diagram as a function of temperature in the light of the finite temperature tricriticality demonstrated in this work. 

\vspace{-0.3cm}
\section{Acknowledgements}

RT, NC, PB and PR thank CeFEMA, an IST research unit whose activities are partially funded by FCT contract UI/DB/04540/2020 for R\&D Units. RT thanks CeFEMA for funding under FCT contract UI/BD/154734/2023 and acknowledges computational resources provided under project 2023.10573.CPCA.A1.

\bibliography{biblio}
\bibliographystyle{apsrev4-2}

\onecolumngrid
\appendix

\section{Derivation of the action}
\label{appa}

 In this appendix we derive the action for the $U(1)$ lattice gauge theory, from the Hamiltonian, given by
 
 \begin{equation}
H =\frac{U}{2} \sum_{r, \mu}\left({n}_{\mu}(r)\right)^{2}-K \sum_{r, \mu<\nu} \cos \left[\Theta_{\mu\nu}(r)\right] \, .
\label{appa:hamiltonian_u1}
\end{equation}
 
 The Hilbert space is spanned by the states $\left| \phi \right\rangle = \otimes_{r, \mu}\left|\phi_{\mu}(r)\right\rangle$ or $|n\rangle=\otimes_{r, \mu}\left|n_{\mu}(r)\right\rangle$ such that $e^{i \hat{\phi}_{\mu}(r)}|\phi\rangle=e^{i \phi_{\mu}(r)}|\phi\rangle$ and $ \hat{n}_{\mu}(r)|n\rangle=n_{\mu}(r)|n\rangle$.

 As the operator $\hat{n}_\mu(r)$ corresponds to the lattice version of the electric field flux through the link $r+\hat{\mu}$, the counterpart of the Gauss law on the lattice is that the charge, $q$, on a lattice site $r$ is given by the sum of the electric flux through the links connected to site $r$, given by the $\hat{n}_\mu(r)$ operators as 
 
 \begin{equation}
     \hat{Q}_{r}=\sum_{\mu}\left[\hat{n}_{\mu}(r)+\hat{n}_{\mu}(r-\hat{\mu})\right]=q \, .
 \end{equation}

We are interested in studying the $\mathrm{U}(1)$ LGT without charges, so we consider the projector to this subspace of the Hilbert space, written as 

\begin{equation}
    P=\prod_{r} \delta_{\hat{Q}_{r}, 0}=\int \prod_{r} \frac{d \theta(r)}{2 \pi} e^{i \sum_{r}\hat{Q}_{r} \theta(r)} \, ,
\end{equation}
which has the properties $P^2 = P$ and $[P,H] = 0$. Thus, the partition function is given by $Z = \operatorname{tr}[e^{-\beta H} P ]$.

In order to numerically simulate this theory, we perform a Trotter decomposition, by separating the partition function $Z$ into $N$ time intervals, with a temporal extent $\Delta \tau = \beta/N$, thus obtaining a partition function of the form $Z =\int D \phi D \theta \sum_{n} e^{-S[\phi, \theta, n]}$ with

\begin{equation}
\begin{aligned}
S[\phi, \theta, n]=&-\Delta \tau K \sum_{\tau, r, \mu<\nu} \cos \left[{\phi}_{\mu}(\tau, r)-{\phi}_{\nu}(\tau, r +\hat{\mu})+{\phi}_{\mu}\left(\tau, r+\hat{\nu}\right)-{\phi}_{\nu}(\tau, r)\right] \\
&-i \sum_{\tau, r, \mu}\left[\phi_{\mu}(\tau, r)-\phi_{\mu}(\tau+1, r)+\theta(\tau, r)+\theta(\tau, r+\hat{\mu})\right] n_{\mu}(\tau, r) \\
&+\Delta \tau \frac{U}{2} \sum_{\tau, r, \mu}\left[n_{\mu}(\tau, r)\right]^{2} \, .
\end{aligned}
\label{isto}
\end{equation}

In order to identify this action with a theory on a $(3 + 1) \mathcal{D}$ lattice, we identify $\hat{\theta}(\tau,r)$ with the phase of a link variable in the temporal direction, $\phi_0(\tau,r)$, so that the the second term in equation \ref{isto} corresponds to the phase of a space-time plaquette.

We can then approximate the sum over $n$ in the partition function in equation \ref{isto} for $\Delta \tau U \gg 1$ using the Villain approximation~\cite{Villain1975}, given by

\begin{equation}
e^{z \cos (\Phi)}=\sum_{n} I_{n}(z) e^{i n \Phi} \simeq \sum_{n} e^{-\frac{1}{2 z} n^{2}+i \Phi n} \, ,
\label{eq:appa:villainapprox1}
\end{equation}
which is valid for $z\gg 1$. Then, we obtain the approximated partition function, given by $Z =\int D \phi   e^{-S[\phi]}$ with the action $S[\phi]$ given in equation \ref{eq:lgt:actionderivedhamiltonian1}, where we relabelled the spacetime coordinates as $n=(\tau,r)$ with directions $\mu, \nu = (0,1,2,3)$ with $\mu = 0$ for the temporal direction.

\section{}
\label{appb}

In this appendix, we describe a systematic perturbative approach to matching the parameters in the action and Hamiltonian formulations of the lattice gauge theory.

A naive matching of the coefficients multiplying the spatial and space-time parts of the action in Eq. \ref{eq:lgt:actionderivedaniisotropic1} and Hamiltonian \ref{hamiltonian_u1} is as follows. We first note that the interval $\Delta \tau$ used in the Trotter decomposition coincides with the lattice spacing in the temporal direction. Then we may write the temperature as $T = \frac{1}{N_t \Delta \tau}$ and find the following relations between the simulation parameters and the Hamiltonian parameters
\begin{equation}
    \frac{K}{U} = \beta^2 \quad \, , \, \quad
    \frac{T}{U} = \frac{\beta\xi}{N_t} \, .
    \label{eq:results:kutu}
\end{equation}

\begin{figure}[h]
    \centering
    \includegraphics[width=0.5\columnwidth]{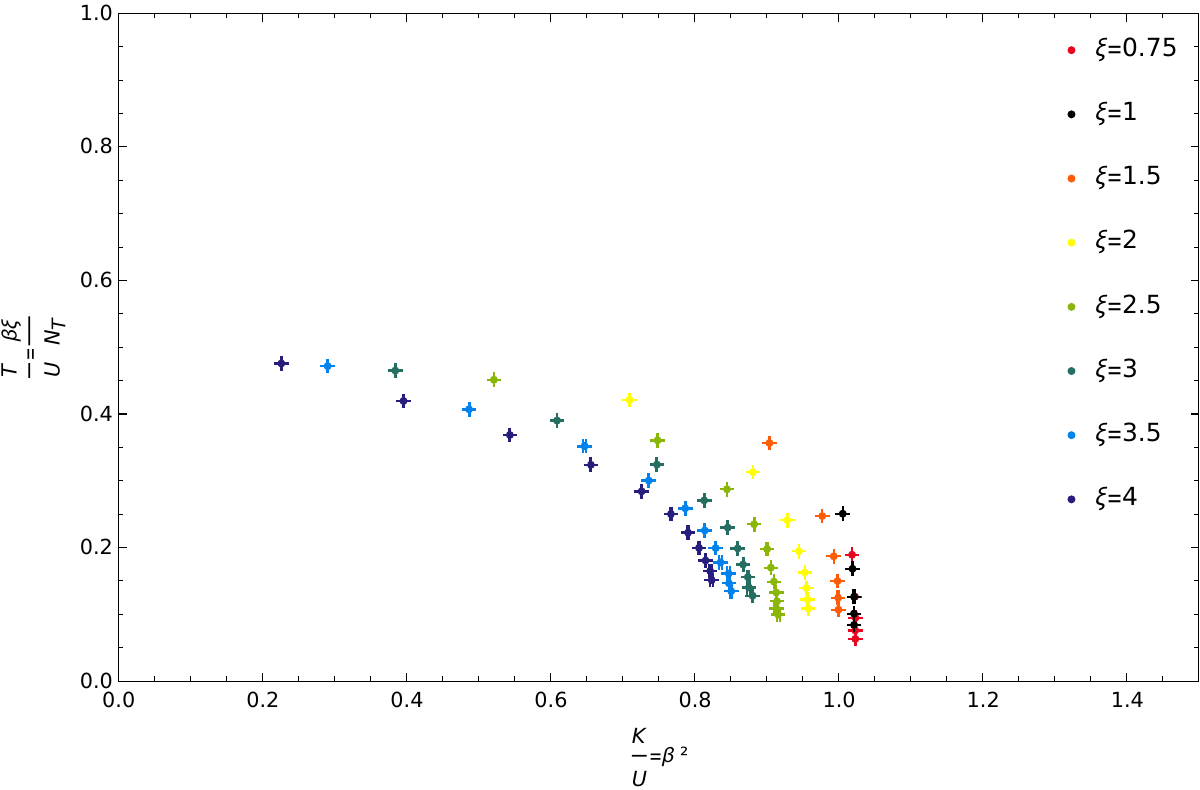}
    \caption{Position of the peaks in the Polyakov loop susceptibility in terms of the quantities in equation \ref{eq:results:kutu}}. 
    \label{fig:resuls:pdraw}
\end{figure}

We determine the critical value of the coupling, $\beta_c$, for different values of $N_t$ and for each value of the anisotropy parameter. Figure \ref{fig:resuls:pdraw} shows these points parameterized using the relations in Eq.~\ref{eq:results:kutu}.

Evidently the points for different $\xi$ in Figure \ref{fig:resuls:pdraw} do not collapse to a single phase transition line. This is a consequence of effectively parameterizing the Hamiltonian from the action at weak coupling (large $\beta$). In particular, as discussed in Appendix \ref{appa}, where we derived the action in equation \ref{isto} from the Hamiltonian, we took $z=\frac{1}{\Delta \tau U}$. However, this approximation is valid only in the limit of large $z$ \cite{Janke1986}. One may systematically improve the correspondence between the two formulations by revisiting the Villain approximation (Eq.~\ref{eq:appa:villainapprox1}) used to derive one from the other.

So instead we take
\begin{equation}
e^{z \cos (\Phi)}=\sum_{n} I_{n}(z) e^{i n \Phi} \simeq \sum_{n} e^{-\frac{1}{2 z_V} n^{2}+i \Phi n} \, ,
\label{eq:results:villain2}
\end{equation}
with $ z_V = - \frac{1}{2\log\left( \frac{I_{1}(z)}{I_{0}(z)} \right)}$. In the limit $z \rightarrow \infty$, we have $z_V \approx z$ and we recover equation \ref{eq:appa:villainapprox1}. 
Using now the approximation in equation \ref{eq:results:villain2} in the action in equation \ref{isto} with $z_V = \frac{1}{\Delta \tau U}$ we obtain

\begin{equation}
\begin{aligned}
S[\phi]=&-\Delta \tau K \sum_{n, \mu<\nu} \cos \left[{\phi}_{\mu}(n)+{\phi}_{\nu}(n +\hat{\mu}) - {\phi}_{\mu}(n  +\hat{\nu}) -{\phi}_{\nu}( n)\right] \\
&- z \sum_{n, \mu}\cos\left[\phi_{\mu}(n) +\phi_0(n+\hat{\mu}) -\phi_{\mu}(n + e_0)-\phi_0(n)\right]  \, ,
\label{eq:results:villainactioncorrection}
\end{aligned}
\end{equation}
with $z$ such that $\frac{1}{\Delta \tau U} = - \frac{1}{2\log\left( \frac{I_{1}(z)}{I_{0}(z)} \right)}$. We match the coefficients multiplying the spatial and space-time parts of the action in equation \ref{eq:results:villainactioncorrection} with the ones in the simulated action, in equation \ref{eq:lgt:actionderivedaniisotropic1}. This leads to
\begin{equation}
    \frac{K}{U} = - \frac{\beta }{\xi }  \frac{1}{2\log\left( \frac{I_{1}(\beta \xi)}{I_{0}(\beta \xi)} \right)} \quad , \quad     \frac{T}{U} = - \frac{1}{ N_t} \frac{1}{2\log\left( \frac{I_{1}(\beta \xi)}{I_{0}(\beta \xi)} \right)} \, .
    \label{eq:results:kutugxbessel}
\end{equation}

\begin{figure}[h]
    \centering
    \includegraphics[width=0.5\columnwidth]{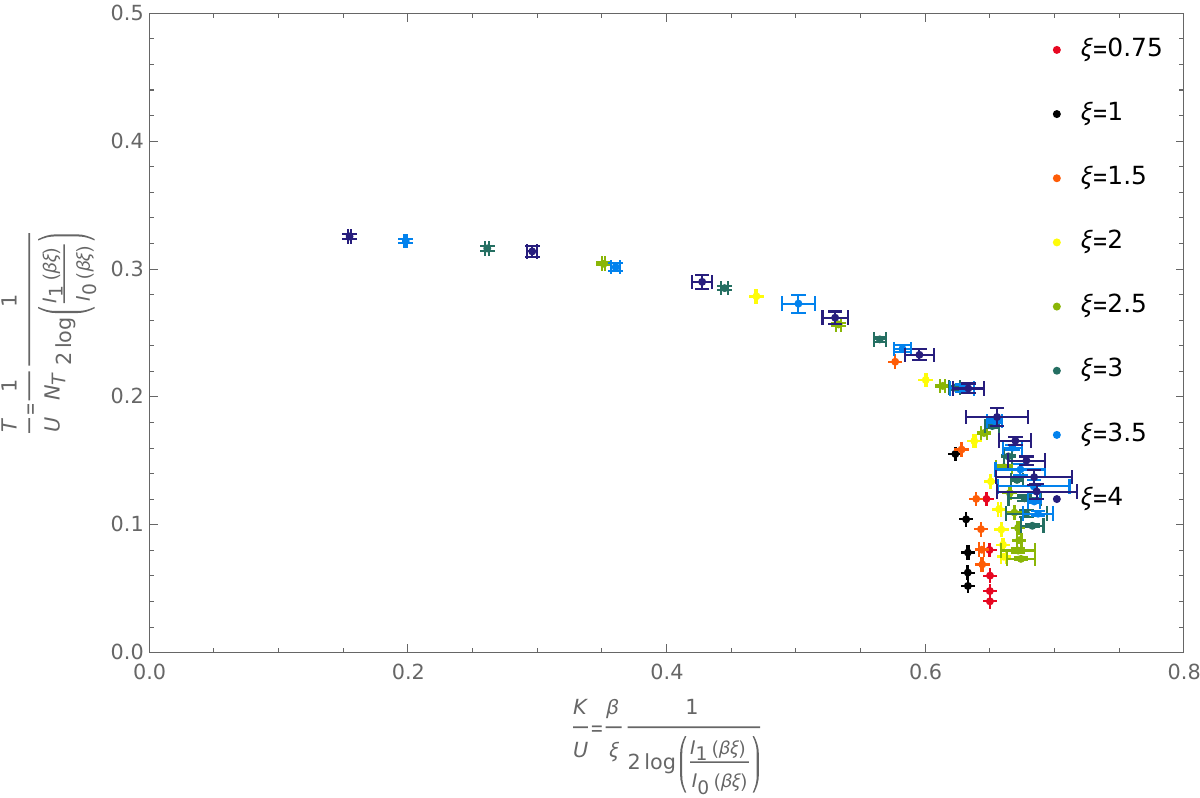}
    \caption{Phase diagram using the relations in equation \ref{eq:results:kutugxbessel}.}
    \label{fig:results:pdvillain}
\end{figure}

We can now plot the critical points at which the phase transition occurs for each value of $N_t$ and $\xi$ in terms of the quantities defined in equation \ref{eq:results:kutugxbessel} and obtain the phase diagram shown in figure \ref{fig:results:pdvillain}.

With this approach, the points for different $\xi$ collapse reasonably well over much of the phase transition line with notable lack of collapse only for large $\frac{K}{U}$. Having understood how to improve the matching conditions directly from the Villain approximation, in the main text we write a general Ansatz for the action that achieves a perfect collapse over the entire phase transition line. 

\section{}
\label{app_table}

\begin{table}[H]
\begin{minipage}{.5\columnwidth}
\centering
\begin{tabular}{c|c|c|c}
$\xi$   & $N_t$ & $\frac{K}{K_0}$   & $\frac{T}{K_0}$    \\ \hline
0.75 & 4  & $0.9956 \pm 0.0001 $& $0.18498 \pm 0.00001 $ \\
0.75 & 6  & $0.9996 \pm 0.0001 $& $0.12356 \pm 0.00001 $ \\
0.75 & 8  & $1.0002 \pm 0.0002 $& $0.09270 \pm 0.00001 $ \\
0.75 & 10 & $1.0000 \pm 0.0002 $& $0.07415 \pm 0.00001 $ \\
0.75 & 12 & $1.0001 \pm 0.0003 $& $0.01797 \pm 0.00001 $ \\ \hline
1    & 4  & $0.9844 \pm 0.0001 $& $ 0.2454 \pm 0.00001 $ \\
1    & 6  & $0.9978 \pm 0.0001 $& $0.16473 \pm 0.00001 $ \\
1    & 8  & $0.9999 \pm 0.00012 $& $0.1237 \pm 0.0001 $ \\
1    & 10 & $0.9999 \pm 0.0005 $& $0.09895 \pm 0.00004 $ \\
1    & 12 & $1.0000 \pm 0.0001 $& $0.08245 \pm 0.00001 $\\ \hline
1.5  & 4  & $0.9009 \pm 0.0001 $& $0.3554 \pm 0.0001 $\\
1.5  & 6  & $0.9776 \pm 0.0001 $& $0.24730 \pm 0.00003 $\\
1.5  & 8  & $0.9944 \pm 0.0001 $& $0.18711 \pm 0.00002 $\\
1.5  & 10 & $0.9999 \pm 0.0002 $& $0.15012 \pm 0.00002 $\\
1.5  & 12 & $1.0009 \pm 0.0015 $& $0.12516 \pm 0.00002 $\\
1.5  & 14 & $1.0014 \pm 0.0005 $& $0.1073 \pm 0.0001 $\\ \hline
2    & 4  & $0.7209 \pm 0.0005 $& $0.4280 \pm 0.0003 $\\
2    & 6  & $0.9107 \pm 0.0005 $& $0.3234 \pm 0.0001 $\\
2    & 8  & $0.9646 \pm 0.0005 $& $0.2503 \pm 0.0001 $\\
2    & 10 & $0.9831 \pm 0.0001 $& $0.2023 \pm 0.00003 $\\
2    & 12 & $0.9923 \pm 0.0006 $& $0.1694 \pm 0.0001 $\\
2    & 14 & $0.9947 \pm 0.0003 $& $0.14541 \pm 0.00004 $\\
2    & 16 & $0.9966 \pm 0.0003 $& $0.12736 \pm 0.00004 $\\
2    & 18 & $0.9975 \pm 0.0010 $& $0.1133 \pm 0.0001 $\\ \hline
2.5  & 4  & $0.5353 \pm 0.0006 $& $0.4634 \pm 0.0005 $\\
2.5  & 6  & $0.7961 \pm 0.0010 $& $0.3834 \pm 0.0004 $\\
2.5  & 8  & $0.9113 \pm 0.0008 $& $0.3098 \pm 0.0003 $\\
2.5  & 10 & $0.9570 \pm 0.0009 $& $0.2547 \pm 0.0002 $\\
2.5  & 12 & $0.9790 \pm 0.0024 $& $0.2149 \pm 0.0005 $\\
2.5  & 14 & $0.9854 \pm 0.0003 $& $0.1849 \pm 0.0001 $\\
2.5  & 16 & $0.9905 \pm 0.0002 $& $0.16225 \pm 0.00002 $\\
2.5  & 18 & $0.9945 \pm 0.0004 $& $0.1445 \pm 0.0001 $\\
2.5  & 20 & $0.9958 \pm 0.0007 $& $0.1302 \pm 0.0001 $\\
2.5  & 22 & $0.9946 \pm 0.0005 $& $0.1183 \pm 0.0006 $\\
2.5  & 24 & $0.9978 \pm 0.0040 $& $0.1086 \pm 0.004 $
\end{tabular}
\end{minipage}
\begin{minipage}{.5\columnwidth}
\centering
\begin{tabular}{c|c|c|c}
$\xi$   & $N_t$ & $\frac{K}{K_0}$   & $\frac{T}{K_0}$    \\ \hline

3    & 4  & $0.3968 \pm 0.0009 $& $0.4800 \pm 0.0009 $\\
3    & 6  & $0.6598 \pm 0.0011 $& $0.4227 \pm 0.0006 $\\
3    & 8  & $0.8316 \pm 0.0016 $& $0.3607 \pm 0.0007 $\\
3    & 10 & $0.9149 \pm 0.0014 $& $0.3044 \pm 0.0005 $\\
3    & 12 & $0.9566 \pm 0.0018 $& $0.2600 \pm 0.0005 $\\
3    & 14 & $0.9745 \pm 0.0010 $& $0.2252 \pm 0.0002 $\\
3    & 16 & $0.9841 \pm 0.0014 $& $0.1981 \pm 0.0003 $\\
3    & 18 & $0.9926 \pm 0.0040 $& $0.1770 \pm 0.0007 $\\
3    & 20 & $0.9944 \pm 0.0050 $& $0.1594 \pm 0.0008 $\\
3    & 22 & $1.0012 \pm 0.0028 $& $0.1455 \pm 0.0004 $\\ \hline
3.5  & 4  & $0.3006 \pm 0.0005 $& $0.4880 \pm 0.0007 $\\
3.5  & 6  & $0.5330 \pm 0.0013 $& $0.4457 \pm 0.0010 $\\
3.5  & 8  & $0.7360 \pm 0.0040 $& $0.4003 \pm 0.0022 $\\
3.5  & 10 & $0.8518 \pm 0.0021 $& $0.3475 \pm 0.0008 $\\
3.5  & 12 & $0.9184 \pm 0.0027 $& $0.3020 \pm 0.0009 $\\
3.5  & 14 & $0.9541 \pm 0.0018 $& $0.2644 \pm 0.0005 $\\
3.5  & 16 & $0.9740 \pm 0.0022 $& $0.2340 \pm 0.0005 $\\
3.5  & 18 & $0.9830 \pm 0.0060 $& $0.2091 \pm 0.0012 $\\
3.5  & 20 & $0.9978 \pm 0.0080 $& $0.1897 \pm 0.0015 $\\
3.5  & 22 & $0.9983 \pm 0.0014 $& $0.1725 \pm 0.0002 $\\
3.5  & 24 & $1.0026 \pm 0.0035 $& $0.1585 \pm 0.0005 $\\ \hline
4    & 4  & $0.2345 \pm 0.0005 $& $0.4929 \pm 0.0010 $\\
4    & 6  & $0.4368 \pm 0.0015 $& $0.4629 \pm 0.0015 $\\
4    & 8  & $0.6257 \pm 0.0024 $& $0.4244 \pm 0.0016 $\\
4    & 10 & $0.7735 \pm 0.0028 $& $0.3821 \pm 0.0014 $\\
4    & 12 & $0.8666 \pm 0.0031 $& $0.3389 \pm 0.0012 $\\
4    & 14 & $0.9201 \pm 0.0033 $& $0.3002 \pm 0.0010 $\\
4    & 16 & $0.9513 \pm 0.0060 $& $0.2674 \pm 0.0018 $\\
4    & 18 & $0.9711 \pm 0.0034 $& $0.2404 \pm 0.0008 $\\
4    & 20 & $0.9830 \pm 0.0040 $& $0.2178 \pm 0.0008 $\\
4    & 22 & $0.9914 \pm 0.0080 $& $0.1989 \pm 0.0016 $\\
4    & 24 & $0.9940 \pm 0.0080 $& $0.1826 \pm 0.0015 $ \\
\end{tabular}
\end{minipage}
\caption{Values of the phase diagram parameters $\frac{T}{K_0}$ and $\frac{K}{K_0}$, calculated for each value of $\xi$ and $N_t$, used in the plot of the phase diagram in figure \ref{fig:results:pdfx}.}
\label{appa:tab1}
\end{table}

\end{document}